\documentclass[11pt]{iopart}
\pdfoutput=1
\usepackage{graphicx}

\newcommand{\be}{\begin{equation}}
\newcommand{\ee}{\end{equation}}
\newcommand{\bea}{\begin{eqnarray}}
\newcommand{\eea}{\end{eqnarray}}

\newcommand{\mt}[1]{\textrm{\tiny #1}}
\def\nc {N_\mt{c}}

\newcommand{\gym}{g_\mt{YM}}

\newcommand{\lqcd}{\Lambda_\mt{QCD}}
\newcommand{\tdiss}{T_\mt{diss}}
\newcommand{\mmes}{M_\mt{mes}}
\newcommand{\vlim}{v_\mt{lim}}
\newcommand{\jpsi}{J/\Psi}

\usepackage{iopams}  
\begin{document}

\begin{flushright}
ICCUB-11-150
\end{flushright}
\vspace{-4mm}

\title[Gauge/string duality applied to heavy ion collisions]{Gauge/string duality applied to heavy ion collisions: Limitations, insights and prospects\footnote{Plenary talk at ``Quark Matter 2011'', Annecy, France.}}

\author{David Mateos}
\vspace{3mm}
\address{Instituci\'o Catalana de Recerca i Estudis Avan\c cats (ICREA), Passeig Llu\'\i s Companys 23, E-08010, Barcelona, Spain}
\vspace{3mm}
\address{Departament de F\'\i sica Fonamental (FFN) \&  Institut de Ci\`encies del Cosmos (ICC), Universitat de Barcelona (UB), Mart\'{\i}  i Franqu\`es 1, E-08028 Barcelona, Spain}

\ead{dmateos@icrea.cat}

\begin{abstract}
Over the last decade
a fruitful interplay has developed between analyses of strongly coupled non-abelian plasmas via the gauge/string duality and the phenomenology of the quark-gluon plasma created in heavy ion collisions. I review the reasons why the gauge/string duality is not a precision tool for QCD physics at present, with emphasis on conceptual issues. I then argue that, nevertheless, the duality can provide valuable insights  at both the quantitive and  the qualitative level. I illustrate this with a few examples, and conclude with a brief discussion of future prospects. 

\end{abstract}


\section{Introduction: The QCD challenge}
Thirty-eight years after the discovery of asymptotic freedom \cite{free}, quantum chromodynamics (QCD) remains a challenge. We have some  tools to analyze it,
but they all have limitations. Perturbation theory requires weak coupling; lattice QCD is a powerful, non-perturbative tool, but it is not well suited for studying real-time phenomena; etc. These limitations have been made particularly apparent by the experimental results at the relativistic heavy ion collider (RHIC) \cite{rhic} and at the large hadron collider (LHC) \cite{lhc}. For our purposes, the most important result is that the quark-gluon plasma (QGP) created in heavy ion collisions  does not behave as a weakly coupled gas of quarks and gluons, but rather as a strongly coupled fluid \cite{fluid}. This severely restricts the applicability of perturbative QCD techniques, and the calculation of many properties of interest requires a real-time formulation beyond the current ability of lattice QCD. 

\begin{figure}
\begin{center}
\includegraphics[scale=1.55]{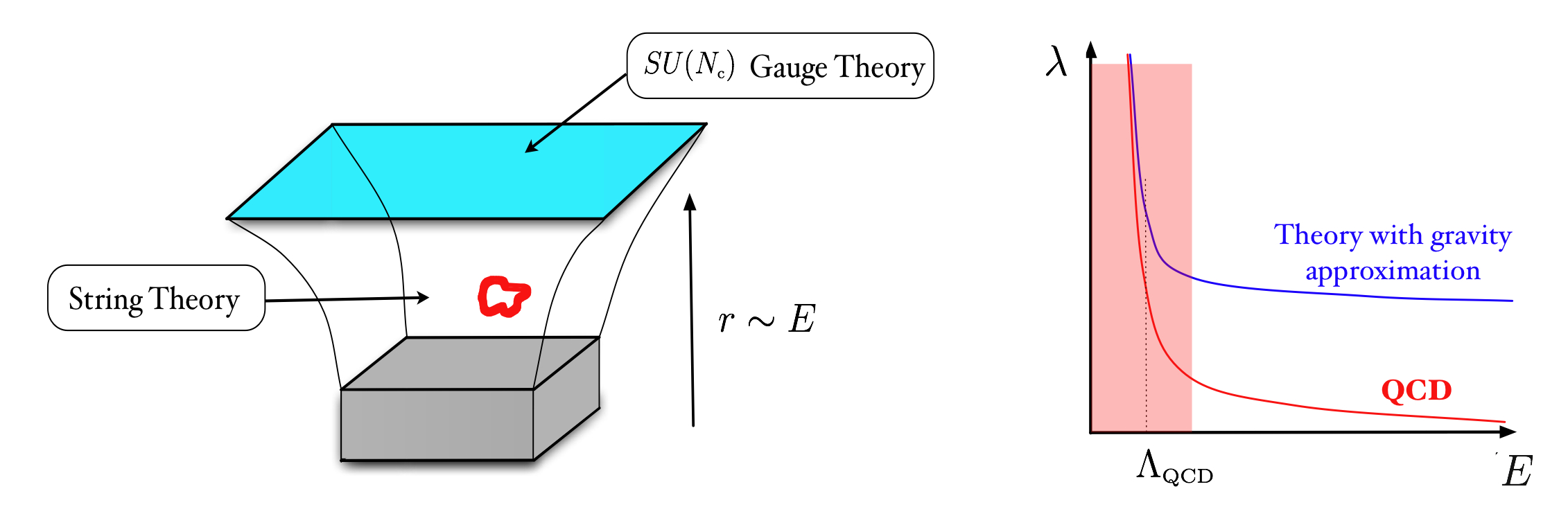}
\caption{(Left) Duality between a gauge theory and a string theory. (Right) Running of the coupling constants.
\label{1}}
\end{center}
\end{figure}

The above difficulties motivate the study of the gauge/string duality \cite{duality} (see \cite{review} for a review of applications to the QGP). This is also known as gauge/gravity duality or AdS/CFT correspondence, but as we will see these are misnomers. The duality  is a conjectured \emph{exact} equivalence between a certain $d$-dimensional $SU(\nc)$ gauge theory and a certain string theory formulated in a $(d+1)$-dimensional spacetime (see figure \ref{1}(left)), valid at all energy scales, for any number of colours $\nc$, and at all values of the `t Hooft coupling constant $\lambda= \gym^2 \nc$. 
This conjecture has not been proven at a mathematical level, but there is ample evidence that it is correct, and we will assume so in this paper. 

Because the gauge theory can be thought of as living at the boundary of the string's spacetime, the duality also goes under the name of `holography'. 
We will refer to the string's spacetime as `the bulk', and to the additional direction in the bulk 
as the holographic radial coordinate, $r$. This coordinate is associated with the renormalization group (RG) or energy scale in the gauge theory, $r \sim E$, in the sense that physical processes occurring at different radial positions in the bulk correspond to processes at different energy scales in the gauge theory. This `geometrization' of the RG flow on the string theory side is one of the most powerful features of the duality.

\section{Limitations}
From the theoretical viewpoint, the discovery of the duality is an extraordinary  achievement. According to the duality, a theory of quantum gravity like string theory is equivalent to an ordinary quantum field theory in flat space, and this is actually the best non-perturbative definition we have at present of a theory of quantum gravity. In addition, the duality has recently found applications to systems other than the quark-gluon plasma as diverse as superfluids, high-$T_c$ superconductors, etc \cite{uni}. In this sense the duality provides a potential framework for a unified description of seemingly very different physical systems. 

In terms of applications to QCD, however, at present the duality suffers from its own limitations, and it must be regarded as a complementary tool. These limitations arise because string theory in a curved spacetime only becomes systematically tractable if two conditions are satisfied. The first condition is that the string coupling constant be arbitrarily weak, $g_s \to 0$, which suppresses quantum corrections on the string side. Since in terms of gauge theory parameters $g_s \sim 1/\nc$, this requires  $\nc \to \infty$. In my opinion this is not the most important limitation. Large-$\nc$ QCD shares many of the fundamental properties of $\nc=3$ QCD: It is asymptotically free, it possesses a dynamically generated scale, it is believed to exhibit confinement and a thermal phase transition into a deconfined phase, etc. Despite differences both at the quantitative level (large-$\nc$ calculations yield just the first term in the $1/\nc$ expansion of real-world QCD) and at the qualitative level (e.g.~deconfinement is a phase transition in large-$\nc$ QCD but only a cross-over in real-world QCD), in my opinion solving large-$\nc$ QCD would constitute fantastic progress. 

Even the classical theory which string theory reduces to in the limit $g_s \to 0$ is still too difficult to admit a systematic treatment. However, this treatment becomes possible if a second condition, $\ell_s / L \to 0$, is satisfied. Here $\ell_s$ and $L$ are the typical size of the string and the typical length scale of the bulk spacetime, respectively. In this limit the string can be approximated by a point particle and the theory reduces to classical general relativity (plus a few extra fields). Incidentally, we see here why `gauge/gravity duality' is a misnomer. The real equivalence is between gauge theory and string theory, and the latter only reduces to  just a gravity theory in a specific limit. 

In terms of gauge theory parameters, the ratio  above is given by $\ell_s / L \sim 1/\lambda^{1/4}$, so we must take $\lambda \to \infty$. In other words, the gauge theory must be strongly coupled at \emph{all} energy scales, which excludes asymptotically free theories ---  see figure \ref{1}(right). At first sight this may not seem like too severe a limitation. After all, the difficult regime of QCD is the strongly coupled infrared (IR) regime, the shaded region in figure \ref{1}(right).  In addition, the gauge theory need not be conformal in order to admit a classical gravity approximation (which is why  `AdS/CFT' is also a misnomer). In fact,  by now many examples are known of gauge theories with a good classical gravity dual that exhibit confinement, spontaneous chiral symmetry breaking, thermal phase transitions, etc. Thus one might be tempted to believe that the lack of asymptotic freedom in these theories is not a crucial obstacle for the purpose of obtaining a good approximation to the IR regime of QCD. However, the reason why in these theories the coupling constant remains strong above $\lqcd$ is that they contain additional degrees of freedom, not present in QCD, with masses precisely of order $M \sim \lqcd$. Consequently, any observable  computed in these theories is `contaminated' by these degrees of freedom. To obtain a reliable approximation to QCD one must take $M\gg \lqcd$. In this decoupling limit the difference between an observable computed in (large-$\nc$) QCD and in a theory with a string dual  becomes of order $\lqcd/M \ll 1$. However, in this limit the gauge coupling  runs precisely as in QCD up to the scale $M$. As a consequence, the condition $\lambda \gg 1$ is violated over a large range of energies and the dual string theory cannot be approximated by classical gravity. 

In a sense that is clear from this discussion, one may say that the reason why QCD is hard to solve is not just that it is strongly coupled in the IR, but that it is strongly coupled in the IR \emph{and} weakly coupled in the UV.

\section{Insights}
Although the two limitations above are not fundamental obstacles but technical difficulties, they mean that, at present, the gauge/string duality is not a precision tool for QCD physics. Nevertheless, certain results from the duality may be universal enough to apply to QCD in certain regimes. In this section I will illustrate this with a few quantitative and qualitative examples.

Possibly the best example of a quantitative result is the calculation of the ratio of shear viscosity to entropy density. This ratio is universal in the sense that 
$\eta/s=1/4\pi$  in any theory with a string dual in the limit $\nc\to \infty$, $\lambda \to \infty$ \cite{ratio,ratio2}. This broad class includes theories in various dimensions, with or without quarks, with or without chemical potentials, etc.  In addition, $1/4\pi$ is in the right ballpark in the sense that results from RHIC and LHC indicate \cite{viscosity} that $1 \lesssim 4\pi \eta/s \lesssim 2.5$
for the QGP at temperatures $T_c \lesssim T \lesssim 2T_c$, where $T_c$ is the deconfinement temperature. Thus, for very good reasons, the calculation of $\eta/s$ in the gauge/string duality is regarded as a benchmark calculation, to the extent that results for this ratio in the heavy ion community are often quoted in units of $1/4\pi$. What I would like to emphasize here is that these reasons do not include the fact that this calculation provides a \emph{precision} prediction for the QGP, since as such it could clearly be off by 250\%. Instead, the reasons why this calculation is important are (i) that it teaches us that  $\eta/s$ \emph{can} be `small' in a strongly coupled QGP, and (ii) that it tells us what \emph{small} means. As a side remark, note that the gauge/string duality does \emph{not} predict that $1/4\pi$ is  a universal lower bound for $\eta/s$. This was conjectured in \cite{ratio2}, but subsequent studies revealed that finite-$\nc$ and/or finite-$\lambda$ corrections can make $\eta/s < 1/4\pi$ --- see \cite{nobound} and references therein.

\begin{figure}
\begin{center}
\includegraphics[scale=1.7]{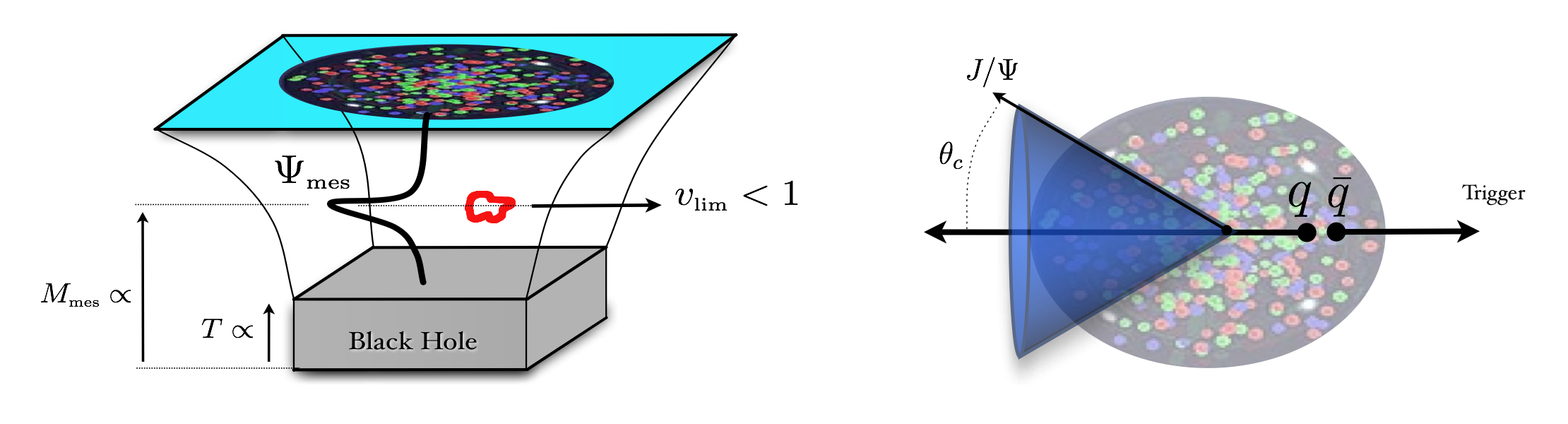}
\caption{(Left) A gauge theory in the deconfined phase is described by a string theory in a black hole background. (Right) Cherenkov emission of heavy quarkonium mesons by a highly energetic quark.
\label{2}}
\end{center}
\end{figure}

The above example illustrates the fact that we should seek results based on universal properties of the duality. For our purposes, the most universal such property is that the deconfined phase of a gauge theory is described by a string theory in a black hole background \cite{witten} --- see figure \ref{2}(left).
This is actually the reason why the ratio $\eta/s$ is universal, since both $\eta$ and $s$ turn out to be proportional to the horizon area. In the rest of this section I will  discuss a few qualitative results related to quarkonium physics (as appropriate for this session) that follow immediately from figure \ref{2}(left). 

The first result is that a sufficiently heavy quarkonium mesons remains bound in the deconfined phase up to some temperature $\tdiss > T_c$ \cite{survival,survival2}. This is due to the fact that the meson wave-function is peaked at a radial position $r_\mt{mes} \propto \mmes$, whereas the position of the black hole horizon is proportional to the gauge theory temperature, $r_\mt{hor} \propto T$ --- see  figure 2(left). This is intuitive in view of the relation between the radial position in the bulk and the energy scale in the gauge theory. It follows that the meson wave-function is essentially unaffected even at $T > T_c$ provided $\mmes$ is sufficiently larger than $T$. If the temperature is further increased to some $\tdiss$ such that $r_\mt{hor} \sim r_\mt{mes}$ then the meson dissociates; in the classical gravity limit this takes place via a first-order phase transition, but this is probably an artifact of the infnite-$\nc$, infinite-coupling approximation. In a specific model one finds \cite{survival2} for the $J/\Psi$ that $\tdiss \sim (1.6-2.1)T_c$. This may be a reasonable ballpark estimate, but I  emphasize that the qualitative prediction that sufficiently heavy quarkonium survives deconfinement is universal within the class of theories with a classical gravity dual, whereas  the precise value of $\tdiss$ is model-dependent. 

A second  consequence of figure 2(left) is the existence of a subluminal limiting velocity for heavy quarkonium mesons moving through the plasma, $\vlim <1$, as discovered in \cite{survival2} and subsequently elaborated upon in \cite{mit}. On the gravity side this limiting velocity is simply the local speed of light at  the position  $r=r_\mt{mes}$, which is lower than the limiting velocity in the vacuum, 
$v_\mt{vac}=1$, because of the redshift caused by the horizon. Typically $\vlim$ can be as low as $0.27 \lesssim \vlim$ near $\tdiss$.

The existence of $\vlim <1$  implies that a highly energetic quark moving through the plasma with velocity $v> \vlim$ can loose energy simply by Cherenkov-radiating heavy quarkonium mesons \cite{che}, as illustrated in figure 2(right). This mechanism is analogous to the radiation of photons by a fast electron moving with $v>v_\mt{light}$ in a dielectric medium, where $v_\mt{light}<1$ is the velocity of light \emph{in} the medium. In the case of a quark in the QGP it turns out that the Cherenkov angle can be quite large, 
$\theta_c =\arccos (\vlim/v_\mt{quark}) \lesssim 74^\circ$. Also, for a temperature range of $T = 200 - 400$ MeV and $\nc = 3$ one finds the ballpark estimate $dE/dx \approx 2 - 8$ GeV/fm, which is comparable in magnitude to other mechanisms of energy loss. If this mechanism is present in the real QGP, then even if it turns out to be subdominant in magnitude its characteristic velocity dependence and geometric properties might make it identifiable. Experimentally, one way to search for this effect would be to trigger on a high-energy jet and look for a statistical excess of heavy quarkonium mesons ($J/\Psi$ mesons, say) on the opposite hemisphere, as illustrated in figure 2(right).

Figure 3(left) shows the dispersion relation for a heavy quarkonium vector meson.
\begin{figure}
\begin{center}
\includegraphics[scale=1.7]{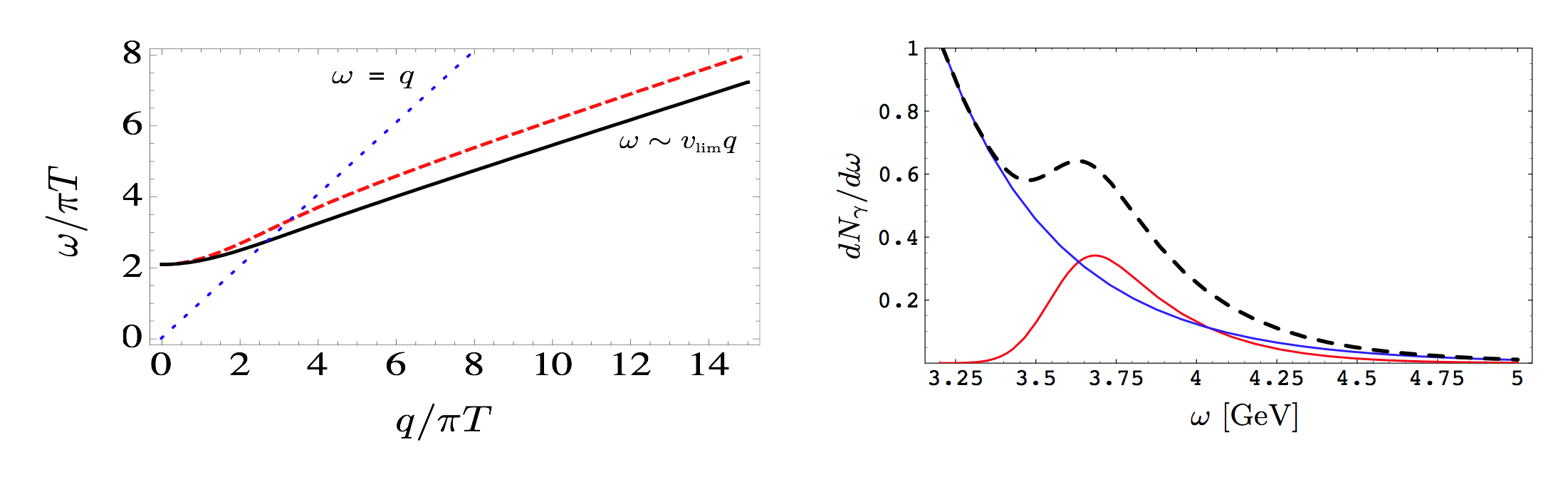}
\caption{(Left) In-medium dispersion relation for the transverse (black, solid curve) and longitudinal (red, dashed curve) modes of a heavy quarkonium vector meson \cite{che}. The straight, blue, dotted line is the lightcone. (Right) Thermal photon spectrum \cite{peak}. The (arbitrary)
normalization is the same for all curves. The continuous,
monotonically decreasing, blue curve is the background from
light quarks. The continuous, red curve is the signal from $\jpsi$
mesons. The dashed, black curve is the sum of the two.
\label{3}}
\end{center}
\end{figure}
Its qualitative form is universal for all mesons in the plasma and follows solely from the information we have at hand: At zero spatial three-momentum, $q=0$, there is mass gap corresponding to the in-medium thermal mass of the mesons, whereas at asymptotically high three-momentum the dispersion relation becomes linear, $\omega \sim \vlim q$, with $\vlim <1$. By continuity, it follows that the meson dispersion relation always crosses the lightcone $\omega=q$ \cite{bright}. The crossing point therefore describes a meson in motion in the plasma with exactly the same kinematical quantum numbers as a photon, i.e.~with lightlike four-momentum. This would be impossible for a massive meson in the vacuum, but in the plasma Lorentz invariance is broken (since the plasma selects a  frame in which it is at rest) and a dramatic modification of the meson dispersion relation is possible. It would be very interesting to investigate these modifications of the dispersion relation in lattice QCD \cite{lattice}. At the crossing point, an in-medium flavourless vector meson like the $\jpsi$ can thus decay into a photon, and this yields a resonance peak in the spectrum of thermal photons produced by the plasma \cite{peak}. In order to estimate whether this could give an observable signal in a real heavy ion collision, the spacetime dependence of the plasma and the in-medium physics of the $\jpsi$ must be modeled. The result for LHC energies, assuming 
$\tdiss (\jpsi)=1.25 \,T_c$, and using a very simple model is shown in figure 3(right) \cite{peak}. I emphasize that whether one obtains a visible peak, simply a statistical enhancement or  an unobservable effect depends sensitively on many parameters related to the  in-medium $\jpsi$ physics. The latter is not sufficiently well understood to make a precise prediction, so all one should take away from  figure 3(right) is that  there could be an observable effect for \emph{some} values of the parameters within the acceptable range. 

A final consequence of figure 2(left) arises by considering the long-wavelength dynamics of perturbations around the equilibrium state on both sides of the duality. On the gauge theory side we expect on general grounds that the long-wavelength dynamics of the plasma must be described by hydrodynamics, so the same must be true on the gravity side. This is indeed the case, as shown at the non-linear level in \cite{fluidgravity}. These authors considered fluctuations around the equilibrium state of the black hole. Working in a derivative expansion, as appropriate in the long-wavelength limit, they showed that Einstein's equations are exactly equivalent, order by order in the expansion, to the hydrodynamic equations of the plasma. In other words, the theory governing the long-wavelength fluctuations of a black hole horizon (in an asymptotically AdS space) is nothing but a hydrodynamical theory whose  transport coefficients at any order can be read off (at least in principle) from the gravity side. By this method the complete set of second-order coefficients for ${\cal N}=4$ super Yang-Mills (SYM) theory was calculated in \cite{fluidgravity,second}.

\section{Prospects}
Because of lack of space I will only briefly mention one direction in which I think progress is likely to take place in the following years. We saw in the previous section that the near-equilibrium dynamics of a gauge theory plasma is equivalent to the near-equilibrium dynamics of a black hole horizon to all orders in the derivative expansion. It turns out that this correspondence can be pushed into the far-from-equilbrium regime. This is extremely difficult to study on the gauge theory side, whereas on the gravity side it corresponds to studying the full-fledged, non-linear but \emph{classical} Einstein equations, generically in the presence of time dependence. This is not an easy task either, but it can be done numerically and it is dramatically simpler than the gauge theory quantum mechanical problem. Progress has already occurred on many fronts, and the interested reader may consult the review \cite{non} and references therein. Here I will just highlight a recent result by Chesler and Yaffe \cite{sheets}, who considered the collision of two sheets of energy in strongly coupled  
${\cal N}=4$ SYM theory. Via a numerical simulation on the gravity side they were able to estimate the time it takes after the collision for a hydrodynamical description to apply. Extrapolated to a heavy ion collision experiment this would yield a thermalization time of 0.3 fm. On the one hand, great care must be exercised in interpreting this result for two reasons. First, I have already argued that any numerical result from the gauge/string duality must be taken at most as a ballpark estimate for QCD. Second, it is not at all obvious that the thermalization process in QCD takes place via a strong-coupling mechanism. It may take place via a weak-couling mechanism, or a combination of both. On the other hand, the result above is very interesting because it suggests that, \emph{if} thermalization takes place via a strong-coupling mechanism, then a thermalization time shorter than 1 fm may be rather natural.

\section{Conclusions}
Although at present the gauge/string duality is not a precision tool for QCD physics, it does provide solvable models of strongly coupled gauge theories that share many of the properties of QCD. As I have illustrated with a few examples related to the QGP, in some cases these can provide valuable insights, both at the quantitative and at the qualitative levels, into problems that are otherwise very difficult or simply impossible to analyze by other methods.

\section*{Acknowledgements}
I am grateful to the organizers of ``Quark Matter 2011'' for the  opportunity to address the heavy ion community. My work is supported by 2009-SGR-168, MEC FPA2010-20807-C02-01, MEC FPA2010-20807-C02-02 and CPAN CSD2007-00042 Consolider-Ingenio 2010.

\end{document}